\documentclass[draft]{article}
\usepackage{fullpage}
\usepackage{amsmath}
\usepackage{amsthm}
\usepackage{amssymb}

\newcommand{\email}[1]{\texttt{#1}}

\newcommand{\abs}[1]{\left\vert{#1}\right\vert}
\newcommand{\ket}[1]{\left| #1 \right\rangle}
\newcommand{\bra}[1]{\left\langle #1 \right|}

\newcommand{\Z}{{\mathbb Z}}
\newcommand{\F}{{\mathbb F}}
\newcommand{\C}{{\mathbb C}}
\renewcommand{\P}{\textrm{P}}
\newcommand{\PF}{{\P\F_q}}
\newcommand{\AGL}{\textrm{AGL}(1; q)}
\newcommand{\AGLd}{\textrm{AGL}(d; 2)}
\newcommand{\AGLp}{\textrm{AGL}(1; p)}
\newcommand{\PGL}{\textrm{PGL}(2; q)}
\newcommand{\GL}{\textrm{GL}(2; q)}
\newcommand{\GLd}{\textrm{GL}(d; 2)}
\newcommand{\PSL}{\textrm{PSL}(2; q)}
\newcommand{\SL}{\textrm{SL}(2; q)}
\newcommand{\polylog}{{\rm polylog}}

\DeclareMathOperator{\Tr}{Tr}
\RequirePackage{dsfont}
\newcommand{\id}{\mathds{1}}

\newtheorem{definition}{Definition}

\newtheorem{lemma}{Lemma}

\begin{document}
\title{Finding conjugate stabilizer subgroups in $\PSL$ and related groups}
\author{Aaron Denney \\
        Center for Quantum Information and Control,\\
        Department of Physics and Astronomy,\\
        University of New Mexico \\
        \email{denney@unm.edu}
   \and Cristopher Moore \\
        Center for Quantum Information and Control,\\
        Department of Computer Science,\\
        University of New Mexico\\
        \& Santa Fe Institute\\
        \email{moore@cs.unm.edu}
   \and Alexander Russell \\
        Department of Computer Science and Engineering\\
        University of Connecticut\\
        \email{acr@cse.uconn.edu}}
\maketitle
\begin{abstract}
We reduce a case of the hidden subgroup problem (HSP) in $\SL$,
$\PSL$, and $\PGL$, three related families of finite groups of
Lie type, to efficiently solvable HSPs in the affine group $\AGL$.
These groups act on projective space in an ``almost'' 3-transitive
way, and we use this fact in each group to distinguish conjugates of
its Borel (upper triangular) subgroup, which is also the stabilizer
subgroup of an element of projective space.  Our observation is mainly
group-theoretic, and as such breaks little new ground in quantum
algorithms.  Nonetheless, these appear to be the first positive results
on the HSP in finite simple groups such as $\PSL$.
\end{abstract}

\section{Introduction: hidden subgroup problems}

One of the principal quantum algorithmic paradigms is the use of the
Fourier transform to discover periodicities hidden in a black-box
function $f$ defined on a group.  In the examples relevant to quantum
computing, an oracle function $f$ defined on a group $G$
has ``hidden periodicity'' if there is a ``hidden'' subgroup $H$ of
$G$ so that $f$ is precisely invariant under translation by $H$ or,
equivalently, $f$ is constant on the cosets of $H$ and takes distinct
values on distinct cosets. The \emph{hidden subgroup problem} is the
problem of determining the subgroup $H$ (or, more generally, a short
description of it, such as a generating set) from such a function.

The standard approach is to use the oracle function $f$ to create
\emph{coset states}
\[ \rho_H = \frac{1}{\abs{G}} \sum_{c \in G} \ket{cH} \bra{cH} \]
where 
\[ \ket{cH} = \frac{1}{\sqrt{\abs{H}}} \sum_{h \in H} \ket{ch} \, . \]
Different subgroups yield different coset states, which must then be distinguished by some series of quantum measurements.

For \emph{abelian} subgroups, sampling these states in the Fourier basis
of the group $G$ is sufficient to completely determine a hidden subgroup
in an efficient manner.  For \emph{nonabelian} subgroups, the Fourier
basis takes the form $\{ \ket{\rho,i,j} \}$ where $\rho$ is the name of
an irreducible representation and $i$ and $j$ index a row and column
in a chosen basis.  Although a number of interesting results have been
obtained on
the nonabelian HSP, the groups for which efficient solutions are known
remain woefully few.  Friedl, Ivanyos, Magniez, Santha, and Sen solve a
problem they call the Hidden Translation Problem, and thus generalize
this further to what they call ``smoothly solvable'' groups: these
are solvable groups whose derived series is of constant length and
whose abelian factors are each the direct product of an abelian group
of bounded exponent and one of polynomial size~\cite{FriedlIMSS02}.
Moore, Rockmore, Russell, and Schulman give an efficient algorithm for
the affine groups $\AGLp = \Z_p \rtimes \Z_p^*$, and more generally
$\Z_p \rtimes \Z_q$ where $q = (p-1)/\polylog(p)$.  Bacon, Childs, and
van Dam derive algorithms for the Heisenberg group and other ``nearly
abelian'' groups of the form $A \rtimes Z_p$, where $A$ is abelian, by
showing that the ``Pretty Good Measurement'' is the optimal measurement
for distinguishing the corresponding coset states~\cite{BaconCD05}.
Recently, Ivanyos, Sanselme, and Santha~\cite{IvanyosSS07,IvanyosSS08}
give an efficient algorithm for the HSP in nilpotent groups of class
$2$.

However, for groups of the greatest algorithmic interest, such as the
symmetric group $S_n$ for which solving the HSP would solve
Graph Isomorphism, the hidden subgroup problem appears to be quite
hard.  Moore, Russell, and Schulman showed that the standard approach
of Fourier sampling individual coset states fails~\cite{defies}.
Hallgren et al. showed under very general assumptions that
highly-entangled measurements over many coset states are necessary
in any sufficiently nonabelian group~\cite{highly-entangled}.  For
$S_n$ in particular, Moore, Russell and \'Sniady showed that the
main proposal for an algorithm of this kind, a sieve approach due to
Kuperberg~\cite{Kuperberg05}, cannot succeed~\cite{quantumsieve}.

It is tempting to think that the difficulty of the HSP on the symmetric
group is partly due to the appearance of the alternating group $A_n$
as a subgroup.  For $n \ge 5$, $A_n$ forms one of the families of
nonabelian finite simple groups.  All known algorithmic techniques
for the HSP work by breaking the group down into abelian pieces, as a
semidirect product or through its derived series.  Since simple groups
cannot be broken down this way, it seems that any positive results on
the HSP for simple groups is potentially valuable.

We offer a small advance in this direction.  We show how to efficiently
solve a restricted case of HSP for the family of finite simple groups
$\PSL$, and for two related finite groups of Lie type.  No new quantum
techniques are introduced; instead, we point out a group-theoretic
reduction to a mild extension of a previously solved case of the HSP.
Unfortunately, this reduction only applies to one set of subgroups, and
there is no obvious generalization that covers the other subgroups.  On
the other hand, we show that a similar reduction works in any group
which acts on some set in a sufficiently transitive way, though this is
unhelpful in many obvious cases.

\section{Reduction}

We start with a trivial observation: suppose we have a restricted case
of the hidden subgroup problem where we need to distinguish among a family of
subgroups $H_1, ..., H_t \subset G$.  If there is a subgroup $F$ whose
intersections $K_i = H_i {\cap} F$ are distinct,
then we can reduce the original hidden subgroup problem to the corresponding
one on $F$, consisting of distinguishing among the $K_i$, by restricting
the oracle to $F$, rather than the original domain $G$.

The subgroups in question will be the stabilizers of one or more
elements under a suitably transitive group action.  Recall the following
definitions:

\begin{definition}
A \emph{group action} of a group $G$ on a set $\Omega$ is a homomorphism $\phi$ from $G$ to the group of permutations on $\Omega$.  In other words,
\[
\phi(g_1 g_2)(x) = \phi(g_1)(\phi(g_2)(x)) \, . 
\]
When the group action is understood, we will often write just $g_1(x)$ for
$\phi(g_1)(x)$.  
\end{definition}

\begin{definition}
A \emph{transitive} group action on a set $\Omega$ is one such that
for any $\alpha, \beta \in \Omega$ there is at least one $g \in G$
such that $g(\alpha)=\beta$.  A \emph{$k$-transitive} group action
is one such that any $k$-tuple of distinct elements
$(\alpha_1,\ldots,\alpha_k)$ can be mapped to any $k$-tuple of distinct
elements $(\beta_1,\ldots,\beta_k)$.  That is, given that $\alpha_i = \alpha_j$ and $\beta_i = \beta_j$
only when $i = j$, there is at least one $g$
such that $g(\alpha_i)=\beta_i$ for all $i=1,\ldots,k$
A group is called $k$-transitive
if it has a $k$-transitive group action on some set.
\end{definition}

\begin{definition}
Given an element $\alpha \in \Omega$, the \emph{stabilizer} of $\alpha$ with respect to a given action by a group $G$ is the subgroup $G_\alpha  = \{ g \in G \mid g(\alpha)=\alpha \}$.  Given a subset $S \subseteq \Omega$, the pointwise stabilizer is 
\[ G_S = \{ g \in G \mid \forall \alpha \in S: g(\alpha) = \alpha \} = \bigcap_{\alpha \in S} G_\alpha \, . \]
When $S$ is small we will abuse notation by writing, for instance, $G_\alpha$ or $G_{\alpha,\beta}$.
\end{definition}

Let's consider the case of the HSP where we wish to distinguish the
one-point stabilizers $G_\alpha$ from each other.  If $G$ is transitive,
these are conjugates of each other, since $G_\beta = g G_\alpha
g^{-1}$ for any $g$ such that $g(\alpha)=\beta$.  Conversely, $g
G_\alpha g^{-1} = G_{g(\alpha)}$, so any conjugate of a stabilizer is
a stabilizer.  Similarly, for each $\alpha$, the two-point stabilizers
$G_{\alpha,\beta}$ labeled by $\beta$ are conjugate subgroups in $G_\alpha$.

Now suppose we restrict our queries to the oracle to $G_\alpha$.  We then get a coset
state corresponding to $G_\alpha \cap G_\beta = G_{\alpha,\beta}$: \[
\rho_{G_{\alpha \beta}} = \frac{1}{\abs{G_\alpha}} \sum_{c \in G_\alpha}
\ket{c G_{\alpha,\beta}} \bra{c G_{\alpha,\beta}} \, . \]
This reduces the
problem of distinguishing the one-point stabilizers $G_\beta$, as
subgroups of $G$, to that of distinguishing the two-point stabilizers
$G_{\alpha,\beta}$ as subgroups of $G_\alpha$---a potentially easier
problem.  Note that we can test for the possibility that $\alpha=\beta$ with a
polynomial number of classical queries, since we just need to check that
$f(1)=f(g)$ for a set of $O(\log |G|)$ generators of $G_\alpha$.

Of course, this whole procedure is only useful if $G_{\alpha, \beta}$ are
distinct when $G_{\beta}$ are distinct, or if there are only a
(polynomially) small number of one-point stabilizers corresponding to each
two-point stabilizer.  Below we give sufficient conditions for this
to be true, and use this reduction to give an explicit algorithm for
distinguishing conjugates of the Borel subgroups in some finite groups
of Lie type, including the finite simple groups \PSL.
Using the transitivity of the group action we can bound
the size of these stabilizers relative to each other and to the original
group, and hence show that they are distinct.

\begin{lemma}
\label{lem:size}
Suppose $G$ has a $k$-transitive group action on a set $\Omega$ where $\abs{\Omega}=s$.  Then for any $j \le k$, if $S \subseteq \Omega$ and $\abs{S}=j$, we have
\[
\frac{\abs{G}}{\abs{G_S}} = \frac{s!}{(s-j)!} \, .
\]
In particular, 
\[
\abs{G_\alpha} = \frac{\abs{G}}{s} \, , \;
\abs{G_{\alpha, \beta}} = \frac{\abs{G}}{s(s-1)} \, , \;
\abs{G_{\alpha, \beta, \gamma}} = \frac{\abs{G}}{s(s-1)(s-2)} \, .
\]
\end{lemma}

\begin{proof}
The index of $G_S$ in $G$ is the number of cosets.  There is one coset for each $j$-tuple to which we can map $S$, and since $G$ is $j$-transitive this includes all $s!/(s-j)!$ ordered $j$-tuples.
\end{proof}

For groups that are at least 3-transitive, the following then holds:
the intersection of two subgroups that are single-point stabilizers of
$\Omega$ has size $1/(s-1)$ of both of the subgroups.  The intersection
with a third stabilizer subgroup is $1/(s-2)$ this size again.  In
particular, this means that when subgroups $G_{\beta}$ and $G_{\gamma}$
are distinct, then their intersections
$G_{\alpha} \cap G_{\beta} = G_{\alpha,\beta}$ and
$G_{\alpha} \cap G_{\gamma} = G_{\alpha,\gamma}$ are distinct, because
their intersection
$G_{\alpha,\beta} \cap G_{\alpha,\gamma} = G_{\alpha,\beta,\gamma}$ is
smaller than either.

In fact, we don't need full 3-transitivity for this argument to hold.
The crucial fact we used was that the number of cosets of
$G_{\alpha,\beta,\gamma}$ was greater than $G_{\alpha,\beta}$
or $G_{\alpha,\gamma}$.  Consider the following definition:
\begin{definition}
\label{almost}
A group is \emph{almost $k$-transitive} if there is a constant
$b$ such that $G$ has an action on a set $\Omega$ which is
$(k-1)$-transitive, and such that we can map any $k$-tuple of distinct elements
$(\alpha_1,\ldots,\alpha_k)$ to at least a fraction $b$ of all ordered
$k$-tuples $(\beta_1,\ldots,\beta_k)$ of distinct elements.
\end{definition}

Strictly speaking, there is a different notion of ``almost'' for
different values of $b$.  Obviously, for any group there is some
value of $b$ low enough that this definition applies.  However, by fixing
$b$ and considering a family of groups we still have a useful concept.

As an example, a group action is \emph{$k$-homogeneous} if any set of
points of size $k$ can be mapped (setwise) to any other set of the same
size.  Since this means that any ordered $k$-tuple can be mapped to at
least $1/k!$ of the ordered $k$-tuples, and since all $k$-homogeneous
group actions are $(k-1)$-transitive~\cite{permutationgroups}, a group
with such an action is almost $k$-transitive with $b=1/k!$ (in fact,
with $b=1/k$).

Applying the above argument to almost $3$-transitive groups shows
that the stabilizer of 3 distinct elements is smaller than the stabilizer
of 2 distinct elements by a factor of $(s-2)b$.  So long as $b \geq 1/(s-2)$,
two-point stabilizers of distinct elements will be distinct.  In the group
families we cover, $b=1/2$, and $s$ grows.

\section{Families of transitive groups}

Which families of groups and subgroups have the kind of transitivity
that let us take advantage of this idea?

Unfortunately, not many do.  We can categorize based on \emph{faithful} group
actions, i.e., those that do not map any group element other than the
identity to the trivial action.  Any non-faithful group action corresponds
to a faithful action of a quotient of the group.
Even the requirement of 2-transitivity in faithful
group actions restricts the choices to a few sporadic groups, or one of eight
infinite families~\cite{permutationgroups}: The symmetric group $S_n$, the
alternating group $A_n$, and six different families of groups of Lie
type.

Obviously the symmetric group $S_n$ is $n$-transitive, and the
alternating group $A_n$ is almost $n$-transitive.  However, the size
$n$ of the set these groups act on is only polynomially large (i.e.,
polylogarithmic in the size of the groups) so we can distinguish the
one-point stabilizers with a polynomial number of classical queries.

The other infinite families are finite groups of Lie type which are defined in
terms of matrices over finite fields $\F_q$ subject to some conditions.
These groups have natural actions by matrix multiplication on column
vectors, or on equivalence classes of column vectors.  The actions
of most of these groups are rather complicated to describe; for more
details, see~\cite[\S 7.7]{permutationgroups}.  Of these, two
are 3-transitive: $\PSL$, and $\AGLd$.

There are also a number of sporadic finite groups that are up to
5-transitive, such as the Matthieu groups $M_{11}$, $M_{12}$, $M_{22}$,
$M_{23}$, $M_{24}$, built on finite geometries.  However, an interesting
fact is that if a group action has a threshold of transitivity, then it
contains all permutations, or at least all even ones: for $k > 5$, all
finite groups with a $k$-transitive action on a set of size $n$ must
contain $A_n$~\cite{permutationgroups}.

\section{$\PSL$ and some relatives}

The most interesting family of simple groups with a faithful almost 3-transitive group action is $\PSL$.  To discuss it, consider instead $\GL$, the group of invertible $2 \times 2$ matrices with entries in the finite field $\F_q$, where $q = p^n$ is the power of some prime $p$.  Its elements are of the form
\[
\begin{pmatrix}\alpha & \beta \\ \gamma & \delta \end{pmatrix}
\]
where $\alpha, \beta, \gamma, \delta \in \F_q$, and $\alpha \delta - \beta \gamma \neq 0$.  We will assume that $q$ is odd; some details change when it is a power of $2$, but the basic results still hold.

A little thought reveals that $\abs{\GL} = (q^2-1)(q^2-q) = (q+1)q(q-1)^2$.  The subgroup $\SL$ consists of the matrices with determinant $1$, so $\abs{\SL} = (q+1)q(q-1)$.  If we take the quotient of these groups by the normal subgroup consisting of the scalar matrices, we obtain $\PGL$ and $\PSL$ respectively.  For $\SL$ the only scalar matrices are $\pm \id$, so $\abs{\PSL} = (q+1)q(q-1)/2$.  

$\GL$ and $\SL$ act naturally on nonzero $2$-dimensional vectors.  For $\PGL$ and $\PSL$, we must identify vectors which are scalar multiples.  This identification turns $\F_q^2 - \{0,0\}$ into the projective line $\PF$.  Each element of $\PF$ corresponds to a ``slope'' of a vector: the vector $\begin{pmatrix} x \\ y \end{pmatrix}$ has slope $x/y$, i.e., $xy^{-1}$ if $y \ne 0$ and $\infty$ if $y=0$.  Thus we can think of $\PF$ as $\F_q \cup \{ \infty \}$, and it has $q+1$ elements.

The action of $\PGL$ and $\PSL$ on $\PF$ is given by
\[
\begin{pmatrix}\alpha & \beta \\ \gamma & \delta \end{pmatrix} 
\begin{pmatrix} x \\ y \end{pmatrix} \\
= \begin{pmatrix}
\alpha x + \beta y \\ \gamma x + \delta y
\end{pmatrix} \, . 
\]
This fractional linear transformation is analogous to the M\"obius transformation defined by ${\rm PGL}_2(\C)$:
\[
\begin{pmatrix}\alpha & \beta \\ \gamma & \delta \end{pmatrix} 
\left( \frac{x}{y} \right) 
= \frac{\alpha x + \beta y}{\gamma x + \delta y} \, , 
\]
which can map any 3 points in the complex projective line $\P\C$ (i.e., the complex plane augmented by the point at infinity, or the Riemann sphere) to any other 3 points.  When we replace $\C$ with the finite field $\F_q$, the action of $\PGL$ remains 3-transitive.  The action of $\PSL$ is 2-transitive, but cannot be 3-transitive, since there are half as many elements as there are 3-tuples.  However, $\PSL$ is almost 3-transitive in the sense defined above with $b=1/2$, since $1/2$ of all 3-tuples can be reached. $\SL$ is also almost 3-transitive: from a given tuple, it reaches the same set of tuples as $\PSL$, with each tuple being hit twice.  As a result, this action is obviously not faithful, for the kernel is $\pm \id$.

Let $G=\PGL$, and consider the one-point stabilizer subgroups of
its action on $\PF$.  A natural one is the Borel subgroup $B$ of
upper-triangular matrices.  Such matrices preserve the set of vectors
of the form $\begin{pmatrix} x \\ 0 \end{pmatrix}$, so we can write
$B=G_\infty$.  There are $q+1$ conjugates of $B$, including itself,
one for each element of $\PF$.  For instance, if we conjugate by
the Weyl element $w=\begin{pmatrix} 0 & -1 \\ 1 & 0 \end{pmatrix}$, we get $w B w^{-1} =
G_0$, the subgroup of lower-triangular matrices, which preserves the set
of vectors of the form $\begin{pmatrix} 0 \\ y \end{pmatrix}$.

\section{An efficient algorithm for distinguishing the conjugates of the Borel subgroup}

Now consider the case of the HSP on these groups where the hidden subgroup is one of $B$'s conjugates, or equivalently, one of the one-point stabilizers $G_s$.  As discussed above, we solve this by restricting the oracle to $B$, and distinguishing the two-point stabilizer subgroups $B \cap G_s = G_{s,\infty}$ as subgroups of $B$.  To do this, we need to describe the structure of $B$ explicitly.  For all three families of matrix groups we discuss, namely $\SL$, $\PSL$, and $\PGL$, $B$ is closely related to the affine group.

In $\PGL$ a generic representative of $B$ can be written
$\begin{pmatrix}{\alpha} & {\beta} \\ 0 & 1 \end{pmatrix}$,
$\alpha \neq 0$, so $\abs{B}=q(q-1)$.  This is exactly the affine group
$\AGL \cong \F_q \rtimes \F_q^*$.  To see this, recall that $\AGL$
consists of the set of affine functions on $\F_q$ of the form
$x \mapsto \alpha x + \beta$ under composition.  Now consider $B$'s
action on $\PF - \{\infty\}$, which we (re)identify with $\F_q$.
For $\PGL$, we have
\[
\begin{pmatrix} \alpha & \beta \\ 0 & 1 \end{pmatrix} 
\begin{pmatrix} x \\ 1 \end{pmatrix}
= \begin{pmatrix} \alpha x + \beta \\ 1 \end{pmatrix} \, .
\]
Obviously these elements compose as $\AGL$, so $B \cong \AGL$.  

The cases of $\SL$ and $\PSL$ are more complicated.  The unit
determinant requirement limits $B$ to elements of the form
$\begin{pmatrix}{\alpha} & {\beta} \\ 0 & {\alpha}^{-1} \end{pmatrix}$.
Thus $\abs{B}=q(q-1)$ again in $\SL$.  For $\PSL$ we identify $\alpha$
with $-\alpha$, so $\abs{B}=q(q-1)/2$.

For $\SL$, we can enumerate the elements as:
\[
\begin{pmatrix} \alpha & \alpha^{-1} \beta \\ 0 & \alpha^{-1} \end{pmatrix}\, . 
\]
Composing two such elements gives us:
\[
\begin{pmatrix} \alpha & \alpha^{-1} \beta \\ 0 & \alpha^{-1} \end{pmatrix} 
\begin{pmatrix} \gamma & \gamma^{-1} \delta \\ 0 & \gamma^{-1} \end{pmatrix} 
=
\begin{pmatrix} \alpha \gamma  & \alpha^{-1} \gamma^{-1} \beta + \gamma^{-1} \alpha \delta \\ 0 & \alpha^{-1} \gamma^{-1} \end{pmatrix} 
= \begin{pmatrix} \alpha \gamma  & (\alpha^{-1} \gamma^{-1}) (\beta + \alpha^2 \delta) \\ 0 & \alpha^{-1} \gamma^{-1} \end{pmatrix} \, .
\]

Here, we still have a semidirect product of the groups $\F_q$ and
$\F_q^*$.  Unlike the affine group, where the multiplicative group
acts directly as an automorphism on the additive group by multiplication, it
instead acts ``doubly'' by multiplying twice, analogous to the
``$q$-hedral'' groups in~\cite{affinegroup} (with $q=p/2$, in their
notation).  Finally, $\PSL$ merely
forgets the difference between $\pm \alpha$.  This quotient group of
$\SL$ can also be seen as a subgroup of the affine group that can only
multiply by the square elements.

In all three cases the HSP on $B$ can be solved efficiently using small
generalizations of the algorithms of~\cite{affinegroup}.  We need to
generalize slightly as \cite{affinegroup} deals only with the case of
$\Z_n \ltimes \F_p$ with $p$ prime --- not a prime power $q=p^n$, as here.
The basic methods remain effective, though we construct and analyze a
slightly different final measurement.  The number and size
of the representations remains the same (with $q$ replacing $p$),
and the methods for constructing
Gelf'and-Tsetlin adapted bases are similar.  As this has not been
published in the literature, we describe the details more fully in the
next section, though only what is necessary for our purposes.

\section{Generalizing the affine group to the prime power case}

Although there can be more types of subgroups than the ones covered
in~\cite{affinegroup}, we are only concerned about one particular type
whose analog was covered there:  $H = (a, 0)$ and its conjugates
$H^b = (1, b)H(1,-b)$, stabilizing the finite field element $b$.
The representation theory is analogous, with $q-1$ one-dimensional
representations (characters) depending only on $a$.  As in the
prime case, we have $q$ conjugacy classes: the identity, all pure
translations, and each multiplication by a different $a$, combined
with all translations.  This leaves us with one $(q-1)$ dimensional
representation, $\rho$.

In the prime case we had:
\[
\rho((a,b))_{j,k} = \begin{cases} \omega_p^{bj} & k = aj \\ 0 &
    \hbox{otherwise} \\ \end{cases} (j, k \in \F_q, \neq 0) \, .
\]
where $\omega_p = \exp (2 \pi i / p)$.  The roots of unity are the
non-trivial additive characters of $\F_p$, indexed by $j$, evaluated at
$b$.  We can extend this to the prime power case simply by replacing
$bj$, with $b \cdot j$
\[
b \cdot j = \Tr bj = \Tr_{\F_{p^n}/\F_p} bj = \sum_{m=1}^n (bj)^{p^m} \, ,
\]
which as $b$ varies, exactly covers the full set of non-trivial linear operators
from $\F_{p^n}$ to $\F_p$, and $\omega_p^{b\cdot j}$ exactly covers the set
of additive characters.
Performing weak measurement on the coset state yields $\rho$ with probability $P(\rho) = 1 - 1/q$.
Conditioned on that outcome, we get the following projection operator:
\begin{align*}
\pi_{H^b}(\rho)_{j,k} = \frac{1}{q-1} \, \omega_p^{b\cdot(j-k)} \, .
\end{align*}

As in~\cite{affinegroup}, we then perform a Fourier transform on the
rows, and ignore the columns.  There they performed the Fourier transform
over $\Z_{p-1}$, as there were $p-1$ rows.  However, the structure for
general $q$ is not $\Z_q^* \equiv \Z_{q-1}$, but $\F_q^*$.  The
interaction we want to capture is the additive one, not
the multiplicative one.  We can still perform the abelian transform
over the additive group $\F_q \equiv \Z_p^n$ --- the zero
component we lack is, of course, zero.  The probability of observing a frequency
$\ell \in \Z_p^n$ is then:

\begin{align*}
P(\ell) &= \abs{\frac{1}{\sqrt{q(q-1)}} \sum_{j \neq 0} \omega_p^{b\cdot j} \omega_p^{-j \cdot \ell}}^2 \\
&= \frac{1}{q(q-1)} \abs{-1 + \sum_{j} \omega_p^{b\cdot j} \omega_p^{-j \cdot \ell}}^2 \\
&= \frac{1}{q(q-1)}\abs{-1 + q \delta_{\ell b}}^2 \\
&= \begin{cases} \frac{1}{q(q-1)} & \ell \neq b \\
   1-\frac{1}{q} & \ell = b \\
   \end{cases} \, .
\end{align*}

For the case of $B$ in $\PSL$, we can analyze the equivalent measurements via the embedding
in the full affine group, just as in the prime case.  
Let $a$ be a generator of the ``even'' multiplicative subgroup of $\F_q^*$,
consisting of elements that are squares.
$H_a^b$ is then elements of the form $(a^t, (1 - a^t)b)$ stabilizing $b$.
For these subgroups, the trivial representation, a ``sign''
representation, and the large representation occur with non-zero probability.  
The first two have vanishingly small probability, $O(1/q)$.

In the following we use the notation $G(m,a) = \sum_{x \in \F_q^*} m(x)
a(x)$ for the Gauss sum of a multiplicative and an additive character, where 
$\chi_{k}(j) = \omega_p^{k \cdot j}$ is the additive character of $\F_q$
with frequency $k \in \Z_p^n$.  We follow the common convention
that non-trivial multiplicative characters vanish at $0$.
We use the quadratic character $\eta$ of $\F_q^*$, which is $1$ for squares,
and $-1$ for non-squares, to select rows and columns which differ
by values in the ``even'' subgroup mentioned above.

Weak measurement gives us the representation $\rho$ with overwhelming probability.  Conditioning on this event, we get the mixed state
\begin{align*}
\rho(H_a^b)_{j,k} &= \frac{\sqrt{2}}{q-1} \sum_{t=1}^{q-1/2} \omega_p^{(1-a^t b)\cdot j} \delta_{k,a^tj} \\
    &= \frac{\sqrt{2}}{q-1} \sum_{t=1}^{q-1/2} \omega_p^{b\cdot (j-k)} \delta_{k,a^tj} \\
    &= \frac{\sqrt{2}}{q-1} \, \omega_p^{b\cdot (j-k)} (1+\eta(jk))/2 \, . 
\end{align*}
Measuring the column $k$ gives us, up to a phase, $\rho(b)_j=\sqrt{\frac{2}{q-1}} \, \omega_p^{b\cdot j} (1\pm\eta(j))/2$.

We again include the zero component, with zero weight, and perform the abelian Fourier transform 
over the additive group $\F_q \equiv \Z_p^n$.  The probability of measuring frequency $\ell$ is
\begin{align*}
P(\ell) &= \frac{1}{q} \abs{\sum_{j} \omega_p^{j \cdot \ell} \rho(b)_{j} }^2 \\
    &= \frac{2}{q(q-1)} \abs{\sum_{j \neq 0} \omega_p^{(b-\ell)\cdot j} (1 \pm \eta(j))/2}^2 \\
    &= \frac{2}{q(q-1)} \abs{\sum_{j \neq 0} \chi_{b-\ell}(j) \pm \sum_{j \neq 0} \chi_{b-\ell}(j)\eta(j) }^2 \\
    &= \frac{1}{2q(q-1)} \abs{ G(1,\chi_{b-\ell}) \pm G(\eta, \chi_{b-\ell})         }^2 \\
    &= \frac{1}{2q(q-1)} \abs{ q \delta_{b,\ell} - 1 \pm \eta(b-\ell)G(\eta, \chi_1) }^2 \\
    &= \frac{1}{2q(q-1)} \abs{ q \delta_{b,\ell} - 1 \pm \eta(b-\ell) i^d q^{1/2}    }^2 \,
\end{align*}
where $d$ is odd for odd $n$ if $p^n \equiv 3 \pmod 4$, and $d$ is even otherwise.

%The probabilities are given by the absolute value squared, evaluated at each $l$.
For $\ell = b$ we have $P(\ell) = (q-1)^2/2q(q-1) = (q-1)/2q$.  For $\ell \neq b$ we have $P(\ell) = (q+1)/4q(q-1)$ if $d$ is odd.
If $\ell \neq b$ and $d$ is even, we have $P(\ell) = (q \pm 2q^{1/2}+1)/4q(q-1)$.  In any case, the probability of observing $b$ is
\begin{align*}
P(b) = \frac{q-1}{2q} = \frac{1}{2} - O(1/q) \, ,
\end{align*}
so repeating this measurement will allow us to identify $\ell=b$ with any desired probability.  As $\SL$ is a small extension of $\PSL$, we can handle it similarly, by Theorem 8 in~\cite{affinegroup}.  

\section{$\AGLd$ and its stabilizer subgroups}

An interesting question is whether it is useful to apply this approach to
the other family of 3-transitive groups. This is the $d$-dimensional
affine group $\AGLd$, consisting of functions on $\F_2^d$
of the form $Av+B$, where $A \in {\rm GL}_d(\F_2)$ and $B \in \F_2^d$.
It can be expressed as a block matrix of the form $\begin{pmatrix} A &
B \\ 0 & 1 \end{pmatrix}$.  It is the semidirect product $\GLd \ltimes
\F_2^d$, and hence obviously not simple.  That it is triply transitive
can be seen by realizing that the affine geometry it acts on has no
three points that are collinear.

The stabilizer subgroups are $2^d$ conjugate subgroups of the original
$\GLd$.  Obviously this stabilizes the point $0$, and is the
largest subgroup that will, as $\GLd$ has two orbits: the zero vector,
and all others.  A general point $P$ is stabilized by translating it to
0 with the element $(A,B) = (1, P)$, applying any element of $\GLd$, and then translating
back.  To apply our method we need to look at the intersections.

Consider the point $\vec{1}=(0,...,0,1)^T$.  Splitting $A$ into two diagonal
blocks of size $(d-1)\times (d-1)$ and $1\times 1$ and two off-diagonal blocks
of size $(d-1)\times 1$ and $1 \times (d-1)$ allows us to see that $\vec{1}$ is
stabilized by a (transposed) copy of ${\textrm AGL}(d-1; 2)$ living
in $\GLd$.  The last column must be $\vec{1}=(0,...,0,1)^T$ to preserve $\vec{1}$.
The large $(d-1)\times (d-1)$ block must be in $\GLd$ to keep the
the entire transformation invertible, and anything in $\GLd$ will
preserve the first $d-1$ 0 bits of $\vec{1}$.  The rest of the last
row can be arbitrary, resulting in a subgroup isomorphic to
${\textrm AGL}(d-1; 2)$.

As a result, distinguishing the stabilizers of points reduces to
distinguishing conjugates of a smaller transposed copy of the affine
group in the general linear group.  This last reduction does not
immediately yield an efficient new quantum algorithm.

\section{Conclusion}

It is interesting to note that although we can Fourier
sample over $\AGLd$ efficiently~\cite{genericqfft}, we don't know how
to do so in the projective groups.  The fastest known classical Fourier transform
for $\SL$ or $\PSL$ takes $\Theta(q^4 \log q)$ time~\cite{ffsl2},
and the natural quantum adaptation of this takes $\Theta(q \log q)$
time~\cite{genericqfft}.  If $q$ is exponentially large, this is
polynomial, rather than polylogarithmic, in the size of the group.  In
the absence of new techniques for the FFT or QFT, this suggests that
we need to somehow reduce the HSP in $\PSL$ to that in some smaller,
simpler group---which was the original motivation for our work.

We conclude by asking whether our analysis of $\AGLd$ can be extended to give an
efficient algorithm distinguishing its stabilizer subgroups, or whether any of the
other 2-transitive groups have usable ``almost'' 3-transitive actions.

\section{Acknowledgement}
This work was supported by the NSF under grant CCF-0829931, and by the DTO under
contract W911NF-04-R-0009.

\bibliography{quantuminfo}

\begin{thebibliography}{10}

\bibitem{BaconCD05}
Dave Bacon, Andrew~M. Childs, and Wim van Dam.
\newblock From optimal measurement to efficient quantum algorithms for the
  hidden subgroup problem over semidirect product groups.
\newblock In {\em FOCS}, pages 469--478. IEEE Computer Society, 2005.

\bibitem{BernsteinV93}
Ethan Bernstein and Umesh Vazirani.
\newblock Quantum complexity theory (preliminary abstract).
\newblock In {\em Proceedings of the 25th Annual ACM Symposium on the Theory of
  Computing}, pages 11--20. ACM, 1993.

\bibitem{permutationgroups}
John~D. Dixon and Brian Mortimer.
\newblock {\em Permutation Groups}.
\newblock Number 163 in Graduate Texts in Mathematics. Springer-Verlag, New
  York, 1996.

\bibitem{FriedlIMSS02}
Katalin Friedl, {G{\'a}bor} Ivanyos, {Fr{\'e}d{\'e}ric} Magniez, Miklos Santha,
  and Pranab Sen.
\newblock Hidden translation and orbit coset in quantum computing.
\newblock In {\em STOC}, pages 1--9. ACM, 2003.

\bibitem{representationtheory}
William Fulton and Joe Harris.
\newblock {\em Representation Theory: a First Course}.
\newblock Number 129 in Graduate Texts in Mathematics. Springer, New York,
  2004.

\bibitem{highly-entangled}
Sean Hallgren, Cristopher Moore, Martin R{\"o}tteler, Alexander Russell, and
  Pranab Sen.
\newblock Limitations of quantum coset states for graph isomorphism.
\newblock In {\em Proceedings of the 38th Annual ACM Symposium on Theory of
  Computing}, pages 604--617, 2006.

\bibitem{IvanyosSS07}
G{\'a}bor Ivanyos, Luc Sanselme, and Miklos Santha.
\newblock An efficient quantum algorithm for the hidden subgroup problem in
  extraspecial groups.
\newblock In Wolfgang Thomas and Pascal Weil, editors, {\em STACS}, volume 4393
  of {\em Lecture Notes in Computer Science}, pages 586--597. Springer, 2007.

\bibitem{IvanyosSS08}
G{\'a}bor Ivanyos, Luc Sanselme, and Miklos Santha.
\newblock An efficient quantum algorithm for the hidden subgroup problem in
  nil-2 groups.
\newblock In Eduardo~Sany Laber, Claudson~F. Bornstein, Loana~Tito Nogueira,
  and Luerbio Faria, editors, {\em LATIN}, volume 4957 of {\em Lecture Notes in
  Computer Science}, pages 759--771. Springer, 2008.

\bibitem{Kuperberg05}
Greg Kuperberg.
\newblock A subexponential-time quantum algorithm for the dihedral hidden
  subgroup problem.
\newblock {\em SIAM J. Comput.}, 35(1):170--188, 2005.

\bibitem{ffsl2}
John~D. Lafferty and Daniel Rockmore.
\newblock Fast {F}ourier analysis for $sl_2$ over a finite field and related
  numerical experiments.
\newblock {\em Experimental Mathematics}, 1(2):115--139, 1992.

\bibitem{finitefields}
Rudolf Lidl and Harald Niederreiter.
\newblock {\em Finite Fields}.
\newblock Cambridge University Press, 1997.
\newblock Number 20 in Encyclopedia of Mathematics and its Applications.

\bibitem{genericqfft}
Cristopher Moore, Daniel Rockmore, and Alexander Russell.
\newblock Generic quantum {F}ourier transforms.
\newblock In {\em Proceedings of the 15th Annual ACM-SIAM Symposium on Discrete
  Algorithms}, pages 778--787, 2004.

\bibitem{affinegroup}
Cristopher Moore, Daniel Rockmore, Alexander Russell, and Leonard~J. Schulman.
\newblock The power of basis selection in {F}ourier sampling: Hidden subgroup
  problems in affine groups.
\newblock In {\em Proceedings of the 15th Annual ACM-SIAM Symposium on Discrete
  Algorithms}, pages 1113--1122, 2004.

\bibitem{defies}
Cristopher Moore, Alexander Russell, and Leonard Schulman.
\newblock The symmetric group defies {{F}ourier} sampling.
\newblock In {\em Proceedings of the 46th Symposium on Foundations of Computer
  Science}, pages 479--488, 2005.

\bibitem{quantumsieve}
Cristopher Moore, Alexander Russell, and Piotr {\'S}niady.
\newblock On the impossibility of a quantum sieve algorithm for graph
  isomorphism: unconditional results.
\newblock In {\em Proceedings of the 39th Annual ACM Symposium on Theory of
  Computing}, pages 536--545, 2006.

\end{thebibliography}
\bibliographystyle{plain}

\nocite{representationtheory}
\nocite{finitefields}
\nocite{BernsteinV93}
% \nocite{HallgrenRT00}
% \nocite{EttingerH98}
% \nocite{EttingerHK04}
% \nocite{GrigniSVV01}

\end{document}